\documentclass[11pt]{article}
\usepackage[a4paper, margin = 1in]{geometry}
\usepackage[sorting = anyt, maxbibnames=9,maxcitenames=2, backend = bibtex, style=authoryear, doi=false, url=false, dashed=false]{biblatex}
\usepackage[titletoc,page]{appendix}
\renewcommand*\appendixpagename{Appendix}
\renewcommand*\appendixtocname{Appendix}

\renewbibmacro{in:}{}
\usepackage{graphicx}
\usepackage{setspace}
\usepackage{gensymb}
\usepackage{enumerate}
\usepackage{kotex}
\usepackage{abstract,lipsum}\usepackage[colorlinks=true,linkcolor=blue,citecolor=blue,urlcolor=blue]{hyperref}

\usepackage{amsmath}
\usepackage{amssymb}
\usepackage{amsthm}
\usepackage{amsfonts}
\usepackage{upgreek}
\usepackage{mathtools}
\usepackage{mathrsfs}
\usepackage{bm}
\newcommand\defeq{\stackrel{\mathclap{\normalfont\tiny\mbox{def}}}{=}}
\usepackage[symbol]{footmisc}

\usepackage{color}
\usepackage{multirow}
\usepackage{algorithm,algorithmic}
\usepackage{verbatim}

\usepackage{tcolorbox}
\usepackage[english]{babel}
\theoremstyle{remark}
\newtheorem*{remark}{Remark}

\begin{document}


\begin{center}
      \fontsize{19pt}{19pt}\selectfont \textbf{Size-dependent fracture in elastomers: experiments and continuum modeling}
      
      \vspace*{0.25in}
      \fontsize{10pt}{10pt}\selectfont Jaehee Lee$^{\mathrm{a},\mathrm{1}}$, Jeongun Lee$^{\mathrm{a},\mathrm{1}}$, Seounghee Yun$^{\mathrm{b}}$, Sanha Kim$^{\mathrm{b}}$, Howon Lee$^{\mathrm{c}}$, Shawn A. Chester$^{\mathrm{d}}$, Hansohl Cho$^{\mathrm{a},\dagger}$\\
      \vspace*{0.2in}
      \fontsize{9pt}{9pt}\selectfont $^{\mathrm{a}}$ Department of Aerospace Engineering, $^{\mathrm{b}}$ Department of Mechanical Engineering, Korea Advanced Institute of Science and Technology, Daejeon, 34141, Republic of Korea \\ 
      \fontsize{9pt}{9pt}\selectfont $^{\mathrm{c}}$ Department of Mechanical Engineering, Seoul National University, Seoul, 08826, Republic of Korea \\
      \fontsize{9pt}{9pt}\selectfont $^{\mathrm{d}}$ Department of Mechanical \& Industrial Engineering, New Jersey Institute of Technology, Newark, NJ, 07102, USA \\ 
\vspace*{0.1in}
$^{\mathrm{1}}$ These authors contributed equally to this work. \\
$^\dagger$ E-mail: hansohl@kaist.ac.kr
\end{center}

\renewenvironment{abstract}
{\small 
\noindent \rule{\linewidth}{.5pt}\par{\noindent \bfseries \abstractname.}}
{\medskip\noindent \rule{\linewidth}{.5pt}
}

\onehalfspacing
\begin{abstract}
\fontsize{10pt}{10pt}\selectfont
Elastomeric materials display a complicated set of stretchability and fracture properties that strongly depend on the flaw size, which has long been of interest to engineers and materials scientists. Here, we combine experiments and numerical simulations for a comprehensive understanding of the nonlocal, size-dependent features of fracture in elastomers. We show the size-dependent fracture behavior is quantitatively described through a nonlocal continuum model. The key ingredient of the nonlocal model is the use of an intrinsic length scale associated with a finite fracture process zone, which is inferred from experiments. Of particular importance, our experimental and theoretical approach passes the critical set of capturing key aspects of the size-dependent fracture in elastomers. Applications to a wide range of synthetic elastomers that exhibit moderate ($\sim 100$\%) to extreme stretchability ($\sim 1000$\%) are presented, which is also used to demonstrate the applicability of our approach in elastomeric specimens with complex geometries. \\
\end{abstract}

\doublespacing
\section{Introduction}
Elastomers are versatile materials for a variety of engineering and biological applications owing to their remarkable, reversible stretchability. However, flaws such as micro-cavities and cracks inherently contained in elastomeric materials significantly impact their stretchability, as is well described in the classical work by \cite{thomas1955rupture} and \cite{greensmith1963rupture} and in the recent work by \cite{hocine2002fracture} and \cite{pharr2012rupture}. The fracture behavior in elastomeric materials is intrinsically size-dependent in the presence of flaws. Furthermore, the size-dependent fracture response is no longer observed when the flaw size is very small (\cite{chen2017flaw, yang2019polyacrylamide}). An intrinsic length scale often represented by the size of a fracture process zone is strongly associated with the nonlocal nature of fracture in elastomeric materials across a range of length scales. The size-dependent, nonlocal responses manifested near the intrinsic length scale render predictive modeling of such extreme events as damage and failure in elastomeric materials challenging.

Over the past decade, there have been extensive studies of the nonlinear fracture mechanics of elastomeric materials under large deformation (\cite{long2011finite, long2015crack, long2021fracture, creton2016fracture}). Specifically, most of these recent studies have focused on characterizing the fracture process zone around a crack tip, where dissipation mainly occurs. While the fracture process zone in an ideal elastomeric network is assumed to be very small, scaled with the end-to-end length of a polymer chain (\cite{lake1967strength}), the majority of dissipation occurs at a much larger length scale than the molecular scale in a realistic elastomeric network. By considering the substantial dissipation within the macroscopic, finite-sized fracture process zone, more realistic physical pictures of failure in elastomeric networks have been studied experimentally (\cite{yin2021essential, li2021effects}) and theoretically (\cite{long2015crack, qi2018fracture}). Furthermore, \cite{qi2019mapping} and \cite{li2023crack} directly measured the nonlinear deformation field near a crack tip using digital image correlation and quantitatively characterized the fracture process zone by computing the energy release rate and energy dissipation near the crack tip in elastomeric materials. The crack-tip field in the materials was also analytically investigated to underpin the nonlinear nature of crack opening and extension involving very large deformation and non-trivial dissipation mechanisms (\cite{long2015crack, qi2018fracture}). It should also be noted that the fracture process zone has been found to be strongly associated with the size-dependent fracture behavior in elastomeric networks (\cite{chen2017flaw, yang2019polyacrylamide}). More specifically, the work by \cite{chen2017flaw} clearly showed that the intrinsic length scale that corresponds to the size of a fracture process zone is associated with the transition from flaw-sensitive to -insensitive rupture in various elastomeric materials; in their experiments, the macroscopic rupture stretch decreased significantly as the initial crack length increased, especially when the initial crack length (or specimen size) was comparable to the intrinsic length scale characterized for each of the elastomeric materials. However, it was found to become insensitive to the initial crack length when the initial crack length was much smaller than the intrinsic length scale.

Indeed, many natural and synthetic materials exhibit nonlocal, size-dependent mechanical behaviors. Accordingly, numerous continuum theories have been proposed to analyze the nonlocal elastic and plastic responses in crystalline metals (\cite{gurtin2007gradient, fleck1994strain}), shape memory alloys (\cite{qiao2011nonlocal,qiao2016computational}) and granular materials (\cite{kamrin2012nonlocal, henann2016finite}). Nonlocal, gradient-type, continuum damage theories have also been presented to address the size-dependent characteristics of fracture in brittle materials, for which the intrinsic length scale associated with the finite fracture process zone was taken into account (\cite{peerlings1996gradient, de1999coupled, le2003calibration}). Furthermore, the phase-field approach has been developed, rooted in the Griffith fracture criterion, for damage evolution in various brittle materials (\cite{francfort1998revisiting, bourdin2000numerical, miehe2010phase}) over the past two decades. In the phase-field approaches, the intrinsic length scale has also been utilized to describe the diffusive damage zone that numerically regularizes a sharp crack topology and, therefore, enables the modeling of mesh-insensitive crack propagation processes. The phase-field approach has attracted much attention for modeling of damage and fracture in elastomeric materials that undergo large deformations prior to the final failure (\cite{gultekin2016phase, kumar2018fracture, russ2020rupture, talamini2018progressive, lee2023finite}) since the work of Miehe and co-workers (\cite{miehe2014phase}). However, in most of these previous studies of the nonlocal continuum modeling of fracture in brittle materials at small to large strains, the intrinsic length scale $l$ has been used as a numerical regularization parameter over the finite, diffusive damage zone without any consideration of its physical interpretation of the size effect, especially during the fracturing of elastomeric materials.

We present here a novel experimental and theoretical approach to exploring the nonlocal, size-dependent fracture responses in elastomeric materials. More specifically, we show how the size-dependence is captured using the experimentally determined intrinsic length scale and how the phase-field model rooted in the nonlocal, gradient-damage theory is able to describe the size-dependent fracture in various elastomeric materials quantitatively with an array of different geometries. To this end, we experimentally characterize the intrinsic length scale that corresponds to the finite size of a fracture process zone in elastomeric materials. The size-dependent fracture responses are then addressed for elastomeric specimens with various geometries. More specifically, we conducted macroscopic tensile tests of polydimethylsiloxane (PDMS) elastomeric specimens with a wide range of initial crack lengths, and the intrinsic length scale was obtained from the experimental data via a nonlinear fracture mechanics analysis. We then argue that this physically motivated length scale can be directly used in a nonlocal phase-field framework that quantitatively describes the size-dependence of fracture responses in the PDMS samples. Then, we investigate the size-dependent fracture of a photocurable 3D-printed elastomer, TangoPlus \footnote{The commercial name of a 3D-printed elastomer (Connex3 objet260, Stratasys Inc.).}, whose intrinsic length scale is an order of magnitude larger than that of PDMS, from which we further address the role and influence of the intrinsic length scale in sensitivity of the fracture features to a notch length (or specimen size) through experiments and numerical simulations. Furthermore, we make use of the nonlocal phase-field framework with the experimentally identified intrinsic length scale to model the fracture responses in a polyurethane (PU) elastomer and an acrylate elastomer (VHB \footnote{The commercial name of a 3M\texttrademark $\,$ dielectric elastomer.}) that exhibit remarkable stretchability prior to the final failure ($>$ 1000\% in stretch). We then demonstrate that the nonlocal continuum modeling framework is capable of capturing notch-root radius-sensitive or -insensitive fracture features across a wide range of specimen sizes in the materials. Finally, we present a key benefit of the nonlocal continuum model that utilizes the physically motivated length scale by demonstrating the capability of the model to describe a more complicated fracture process involving nucleation, propagation, and merging of cracks and the final failure in an elastomeric specimen with arbitrary, complex geometric features. Overall, throughout this comprehensive, experimental and theoretical investigation, we aim to advance our understanding of the physical principles underlying the nonlocal, size-dependent fracture behaviors in elastomeric materials.

The manuscript is organized as follows. In Section \ref{section2}, we experimentally characterize the intrinsic length scale and demonstrate that the nonlocal model that utilizes the length scale is capable of capturing the main features of the size-dependent fracture behavior in the PDMS and TangoPlus materials. Then, in Section \ref{section3}, we show that the nonlocal model can also account for the size-dependence associated with the notch-root radius effects on fracture. Furthermore, fracture of an elastomeric specimen with a complex geometry is explored in both experiments and numerical simulations. Finally, we summarize the main findings and provide final remarks in Section \ref{section4}. Additionally, the experimental and numerical details are provided in the Appendix, and the associated source codes that generate the numerical results presented throughout the main body of the manuscript are available online in the following GitHub repository, at \href{https://github.com/solidslabkaist/elastomer_fracture}{https://github.com/solidslabkaist/elastomer\_fracture}.

\section{Nonlocal fracture behavior of elastomeric materials}
\label{section2}
\subsection{Experimental characterization of the fracture process zone and size-dependent behavior}
\label{section:Fracture_process_zone}

\begin{figure}[b!]
\centering
\includegraphics[width=1.0\textwidth]{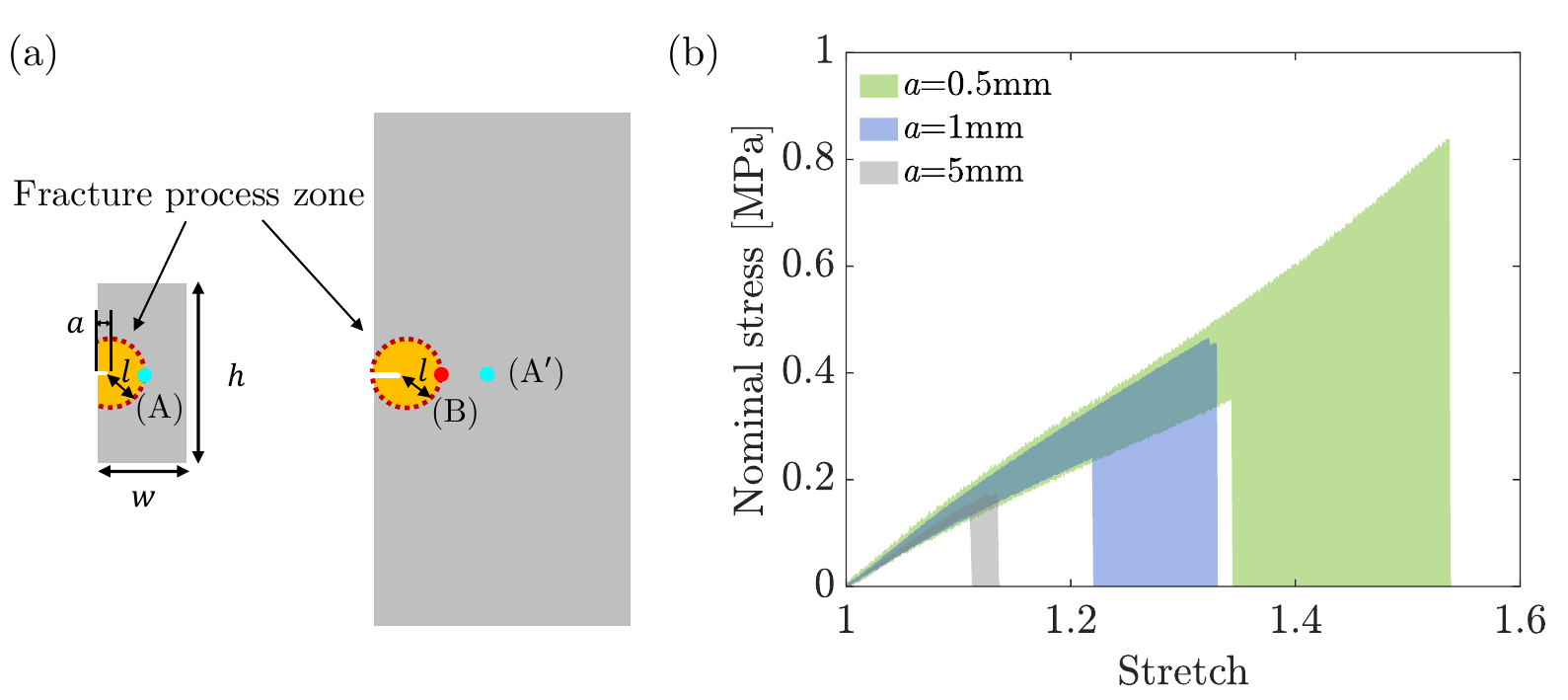}
\vspace{-0.3in}
\caption{(a) Schematic illustration of single-edge-notched specimens with a fracture process zone and (b) nominal stress vs. stretch curves of single-edge-notched PDMS specimens with initial crack lengths of $a$ = \{0.5, 1, 5\}mm in macroscopic tension}
\label{fig:experiment_result}
\end{figure}
First, we present a physical picture of the size-dependent fracture behavior in elastomeric materials in the presence of a fracture process zone near the crack tip. Figure \ref{fig:experiment_result}(a) illustrates the fracture process zone with an intrinsic length scale of $l$ for single-edge-notched specimens of different sizes. From the classical Griffith's fracture criterion (\cite{griffith1921the}), crack propagation occurs when the deformation energy reaches a critical energy density within the fracture process zone. As schematically shown, the fracture process zone characterized by the intrinsic length scale in the smaller specimen occupies a larger portion in the total specimen area compared to the larger specimen. We then compare the deformation energy at two material points (A) and (B), located at a distance of $l$ from the crack tip in the two specimens. Due to geometrical similarities, the relative position of point (A$'$) in the larger specimen is identical to that in point (A) in the smaller specimen; i.e., when we scale down to the smaller specimen, point (A) in the smaller specimen corresponds to point (A$'$) in the larger specimen. The local strain at point (B) is hence larger than that at point (A$'$) in the larger specimen, which is identical to the local strain at point (A) in the smaller specimen under the same level of macroscopic strain. Therefore, the deformation energy due to molecular stretching at this point (here, B) at a distance, $l$, from the crack tip in the larger specimen first reaches the critical energy density under a smaller macroscopic strain level. Note that crack propagation is delayed in the smaller specimen because the deformation energy at the point at a distance, $l$, from the crack tip reaches the critical energy density at a larger macroscopic strain.

Next, we characterize the intrinsic length scale $l$ that accounts for the nonlocal, size-dependent fracture phenomena, following the early studies of size-dependent fracture in brittle materials (\cite{bavzant1997scaling, chen2017flaw, long2021fracture}). From a macroscopic fracture mechanics perspective based on the Griffith-type critical energy release rate criterion (\cite{rivlin1953rupture, thomas1955rupture}), the crack propagates when the energy release rate reaches a critical value, denoted as $\mathit{\Gamma}$, which has the dimension of energy per unit area. At the molecular level, however, the failure criterion for crack propagation in an elastomeric network is described as the chain scission mechanism (\cite{lake1967strength, mao2017rupture}), where they argued that the extension of a crack in an elastomeric network occurs when the internal energy due to stretching in molecular bonds reaches a critical bond dissociation energy. We then define the critical energy density (or work to rupture, \cite{chen2017flaw}), denoted $W^{\ast}$, which corresponds to the total energy per unit volume required for failure within the molecular chain network. Considering the relation between the energy density $W$ due to deformation, the energy release rate $G$ and the distance to the crack tip, $r$, expressed as $W \sim G / r$ that also appears in the nonlinear fracture mechanics approach (\cite{long2015crack}), the energy density $W$ within the fracture process zone approaches $W^{\ast}$ as the energy release rate $G$ reaches a critical value, $\mathit{\Gamma}$ for the extension of a crack. Therefore, the intrinsic length scale $l$ that corresponds to the size of the fracture process zone is determined as the ratio of the microscopic critical energy density $W^{\ast}$ to the macroscopic critical energy release rate $\mathit{\Gamma}$; i.e., the intrinsic length scale, $l = \mathit{\Gamma}/W^{\ast}$ emerges. We now consider the size-dependent fracture behavior of an exemplar elastomeric material, PDMS. We identified the intrinsic length scale $l$ for this elastomeric material by determining the critical energy release rate $\mathit{\Gamma}$ and the critical energy density $W^{\ast}$ from experiments, as detailed in the Appendix \ref{appendix:parameter}. Here, the intrinsic length scale $l$ was estimated to be $\sim$ 0.08mm with the experimentally obtained $\mathit\Gamma \sim 0.25$ mJ/mm$^2$ and $W^{\ast} \sim$ 2.7 mJ/mm$^3$.

\vspace{0.2in}
\begin{remark}
The macroscopic critical energy release rate $\mathit{\Gamma}$ can be decomposed into two contributions, i.e., $\mathit{\Gamma} = \mathit{\Gamma}_0 + \mathit{\Gamma}_\mathrm{D}$, where $\mathit{\Gamma}_0$ represents the intrinsic fracture energy for bond dissociation throughout the elastomeric network and $\mathit{\Gamma}_\mathrm{D}$ represents the bulk energy dissipation through other dissipation mechanisms involving rate-dependent inelasticity and Mullins’ effect. In this work, we assume that the bulk energy dissipation is negligible in the elastomeric network and, therefore, that the intrinsic length scale is related to only the intrinsic fracture energy; i.e., $\mathit{\Gamma} \approx \mathit{\Gamma}_0$.
\end{remark}
\vspace{0.2in}

We then explored the size-dependent fracture behavior of single-edge-notched PDMS specimens with notch lengths of $a=$ \{0.5, 1, 5\} mm. It should be noted that the ratio of the notch length to the specimen size (width and height) was taken to be fixed as the specimen width $w = 10a$ and the specimen height $h = 20a$, and the specimen thickness was 0.5 mm for all cases. The specimen width and height were set to be much larger than the initial notch length in order to avoid boundary effects. Figure \ref{fig:experiment_result}(b) shows the nominal stress-stretch curves up to the rupture stretches of the specimens during macroscopic tensile loading tests. Note that the initial stress-stretch responses are identical in all cases due to geometric similarity. We further note that the initial notch in the specimens was made using a razor blade. As shown, we clearly observed that the macroscopic rupture stretch increased as the notch length (or specimen size) decreased with high sensitivity of the rupture stretch to the notch length. In the following section, we demonstrate that the physically motivated intrinsic length scale, $l$, estimated from the experimental data can be directly used in a predictive, nonlocal continuum modeling framework of the size-dependent fracture responses in the elastomeric materials.

\subsection{Nonlocal continuum model for size-dependent fracture in elastomers}
\label{section:modeling}
A nonlocal continuum modeling framework that incorporates the experimentally obtained intrinsic length scale $l$ for the elastomeric materials is presented.

\paragraph{Kinematics, constitutive theory and governing partial differential equations} \mbox \\

A motion $\bm\upvarphi$ is defined as a one-to-one mapping $\mathbf{x}=\bm\upvarphi(\mathbf{X},t)$ with a material point $\mathbf{X}$ in a fixed undeformed reference and $\mathbf{x}$ in a deformed spatial configuration with deformation gradient $\mathbf{F}\defeq\frac{\partial \bm\upvarphi}{\partial\mathbf{X}}$. Then, we define the following:

\onehalfspace
\begin{equation}
\begin{aligned}
& \text{isochoric part of } \, \mathbf{F} && \bar{\textbf{F}}=J^{-\frac{1}{3}}\textbf{F}, J\defeq\det\textbf{F}\\
& \text{right Cauchy-Green tensor} && \mathbf{C}=\mathbf{F}^\top\mathbf{F} \\
& \text{isochoric right Cauchy-Green tensor} && \Bar{\textbf{C}}=\bar{\mathbf{F}}^{\top}\bar{\mathbf{F}} = J^{-\frac{2}{3}} \mathbf{C} \\
\vspace{0.05in}
\end{aligned}
\end{equation}
\doublespacing

We introduce the scalar damage field $d \in [0, 1]$ which characterizes an intact state by $d=0$ and a fully damaged state by $d=1$. Then, the free energy $\psi_{\mathrm{R}}$ in an undeformed reference degraded due to damage evolution has the form of 
\begin{equation}
\begin{aligned}
\psi_{\mathrm{R}} & = \hat{\psi}_{\mathrm{R}}(\mathbf{F},\lambda_{b},d,\nabla d) \\
& = g(d) \, \hat{\varepsilon}_{\mathrm{R}}^{0} (\mathbf{F}, \lambda_{b}) - \vartheta \, \hat{\eta}_{\mathrm{R}} (\mathbf{F}, \lambda_{b}) + \hat{\psi}_{\mathrm{R,nonlocal}}(\nabla d),
\end{aligned}
\end{equation}
where $\hat{\varepsilon}_{\mathrm{R}}^{0}$ is the internal energy of the undamaged body and $\hat{\eta}_{\mathrm{R}}$ is the configurational entropy of the chain network with an absolute temperature of $\vartheta$. Note that only the internal energy part out of the total free energy degrades with damage evolution driven by the internal energy change due to bond-stretching (\cite{lake1967strength, mao2017rupture}). The degradation function takes the form of $g(d)=(1-d)^{2}$. Following the nonlocal continuum theory, a spatial derivative of the damage variable in the reference configuration, $\nabla d = \frac{\partial d}{\partial \mathbf{X}}$, is involved in the free energy function to account for the nonlocal, size-dependent fracture behavior. The nonlocal contribution of the free energy is defined as
\begin{equation}
\begin{aligned}
\hat{\psi}_{\mathrm{R,nonlocal}}(\nabla d)=\frac{1}{2}\varepsilon^{f}_{\mathrm{R}} l^{2} |\nabla d| ^{2},
\end{aligned}
\end{equation}
where $\varepsilon^{f}_{\mathrm{R}}$ is the bond dissociation energy per unit reference volume. Note that the damage variable evolves within the intrinsic length scale $l$, which represents the size of the fracture process zone obtained from our experimental results.

Here, we employ the effective bond stretch $\lambda_{b}={L_{t}}/{L_{0}}$, where $L_{0}$ is the initial Kuhn segment length and $L_{t}$ is the current Kuhn segment length, to describe the energetic process during rupture in an elastomeric network (\cite{mao2017rupture, talamini2018progressive}). Upon a bond stretch, the undamaged internal energy is defined as
\begin{equation}
\hat{\varepsilon}_{\mathrm{R}}^{0} = \frac{1}{2}NnE_{b}\left(\lambda_{b} - 1 \right)^{2} + \frac{K}{2}(J-1)^{2},
\end{equation}
where $N$ is the number of chains per unit volume, $n$ is the number of Kuhn segments in a chain, $E_{b}$ is the bond stiffness related to the stretching of Kuhn segments, and $K$ is the bulk modulus. The configurational entropy of the chain network is expressed with the nearly incompressible Arruda-Boyce representation (\cite{arruda1993three}), by taking the form of
\begin{equation}
\begin{aligned}
\hat{\eta}_{\, \mathrm{R}} & = - Nnk_{b} \left[ \left(\frac{\bar{\lambda}\lambda^{-1}_{b}}{\sqrt{n}} \right) \beta + \ln\left( \frac{\beta}{\sinh \beta} \right) \right] \quad \text{where} \quad \beta = \mathcal{L}^{-1} \left(\frac{\bar{\lambda}\lambda^{-1}_{b}}{\sqrt{n}} \right),
\label{eq:entropyAB}
\end{aligned}
\end{equation}
with the average stretch expressed as $\bar{\lambda}=\sqrt{\frac{1}{3}\text{tr}\bar{\mathbf{C}}}$, Boltzmann's constant $k_{b}$, and where $\mathcal{L}^{-1}$ denotes the inverse of the Langevin function $\mathcal{L}(x) = \coth x \, - \, x^{-1}$. The degraded Piola stress in the modified Arruda-Boyce representation with bond-stretching can be expressed by

\begin{equation}
\begin{aligned}
\mathbf{T}_{\mathrm{R}} & = \bar{\mu} \left(J^{-2/3} \mathbf{F} - \bar{\lambda}^{2} \mathbf{F}^{-\top} \right) + \left(1-d \right)^{2} K (J-1) J \mathbf{F}^{-\top} \\
\text{where} \, \, \bar{\mu} & = \dfrac{N k_{b} \vartheta}{3} \dfrac{\sqrt{n}}{\bar{\lambda} \lambda_{b}} \mathcal{L}^{-1}\left(\dfrac{\bar{\lambda} \lambda_{b}^{-1}}{\sqrt{n}} \right).
\label{eq:stressAB}
\end{aligned}
\end{equation}

Furthermore, the bond stretch, $\lambda_{b}$, is determined by solving the following implicit equation: 
\begin{equation}
(1-d)^{2} \, E_{b}(\lambda_{b} - 1) - k_{b} \vartheta \left( \frac{\bar\lambda}{\sqrt{n} \lambda^{2}_{b}}\right) \mathcal{L}^{-1} \left( \frac{\bar{\lambda} \lambda_{b}^{-1}}{\sqrt{n}} \right) = 0.
\label{eq:bond-stretch}
\end{equation}
\noindent
This implicit equation is derived from the principle of virtual power, in which the bond stretch minimizes the free energy, i.e., $\frac{\partial \hat{\psi}_{\mathrm{R}}}{\partial \lambda_{b}}=0$. 

For the coupled gradient-damage theory, the governing partial differential equations for macroforce balance (\cite{gurtin2010mechanics}) and microforce balance (\cite{talamini2018progressive,lee2023finite}) in the undeformed reference are given as,
\begin{equation}
\begin{aligned}
\begin{rcases}
& \mathrm{Div} \mathbf{T}_\mathrm{R} = \mathbf{0} \quad \text{in} \quad \mathcal{B}, \\ 
& \zeta_{\mathrm{R}} \dot{d} = 2 (1-d)\mathcal{H}_{\mathrm{R}} - \varepsilon^{f}_{\textrm{R}} (d - l^2 \Delta d) \quad \text{in} \quad \mathcal{B}, \\ 
\end{rcases}
\end{aligned}
\label{eq:strong_form}
\end{equation}
where $\zeta_{\mathrm{R}}$ is a positive-valued parameter associated with the rate-dependent part in microforce balance. For the time interval $t \in [0,T]$, we consider a set of boundary conditions given by $\bm\upvarphi=\check{\bm\upvarphi}$ on $\mathcal{S}_{\bm\upvarphi} \times [0,T]$ and $\mathbf{T}_{\mathrm{R}}\mathbf{n}_{\mathrm{R}}=\check{\mathbf{t}}_{\mathrm{R}}$ on $\mathcal{S}_{\mathbf{t}_{\mathrm{R}}} \times [0,T]$ with complementary subsurfaces of a boundary $\partial\mathcal{B}$ of a body $\mathcal{B}$; i.e., $\partial\mathcal{B}=\mathcal{S}_{\bm\upvarphi} \cup \mathcal{S}_{\mathbf{t}_{\mathrm{R}}}$ and $\mathcal{S}_{\bm\upvarphi} \cap \mathcal{S}_{\mathbf{t}_{\mathrm{R}}} = \varnothing$ for the macroforce balance. Similarly, we take the following boundary conditions, $\dot{d}=0$ on $\mathcal{S}_{d} \times [0,T]$ and $\nabla d \cdot \mathbf{n}_{\mathrm{R}}=0$ on $\mathcal{S}_{\nabla d} \times [0,T]$ with complementary subsurfaces; i.e., $\partial\mathcal{B}=\mathcal{S}_{d} \cup \mathcal{S}_{\nabla d}$ and $\mathcal{S}_{d} \cap \mathcal{S}_{\nabla d} = \varnothing$ for the microforce balance. Furthermore, we define a monotonically increasing history function (\cite{miehe2010phase,talamini2018progressive}),
\begin{equation}
\mathcal{H}_\mathrm{R}(t)=\max\limits_{s \in [0, t]} \langle \hat{\varepsilon}^0_{\mathrm{R}}(\mathbf{F}(s),\lambda_{b}(s)) - \varepsilon^{f}_{\mathrm{R}}/2 \rangle,
\label{eq:historyfunction}
\end{equation}
where $\langle \; \bullet \; \rangle=( \; \bullet \; +| \bullet |)/2$ is the Macaulay bracket that enforces the constraint $d \in [0,1]$ and irreversibility of the damage growth ($\dot{d} > 0$). Here, note that the damage variable, $d$, evolves only when the internal energy $\hat{\varepsilon}^0_{\mathrm{R}}$ due to bond-stretching is greater than $\varepsilon^{f}_{\mathrm{R}}$.

\subsection{Experiment vs. model}
\label{section:exp_vs_model}
\subsubsection{Size-dependent fracture behavior in PDMS}
\label{section:PDMS}
The nonlocal continuum modeling results are presented and compared against corresponding experimental data for the PDMS specimens described in Section \ref{section:Fracture_process_zone}. To this end, the gradient-damage theory utilizing the experimentally identified intrinsic length scale has been numerically implemented within a finite element solver for boundary value problems of the macroscopic tension tests of notched specimens; for numerical details, see our prior work (\cite{lee2023finite}). The material parameters used in the numerical simulations are summarized in Table \ref{Tab:pdms}. Detailed identification procedures for the material parameters are presented in the Appendix \ref{appendix:parameter}. 
\begin{table}[b!]
\doublespacing
\hspace{-0.1in}
\begin{tabular}{ccccccc}
\hline
 $\mu=N k_{b} \vartheta$ {[}MPa{]} & $K$ {[}MPa{]} & $\quad n \quad$ & $\varepsilon^{f}_{\mathrm{R}}$ {[}mJ/mm$^3${]} & $\bar{E}_b=NnE_{b}$ {[}MPa{]} & $l$ {[}mm{]} & $\zeta_{\mathrm{R}}$ {[}kPa$\,\cdot\,$s{]} \\
\hline
 0.38 & 38 & 1.8 & 1.4 & 30 & 0.08 & 1 \\
\hline
\end{tabular}
\vspace{-0.2in}
\caption{Material parameters used in the numerical simulations with the PDMS material.}
\label{Tab:pdms}
\end{table}

\begin{figure}[p!]
\centering
\vspace{-0.2in}
\includegraphics[width=1.0\textwidth]{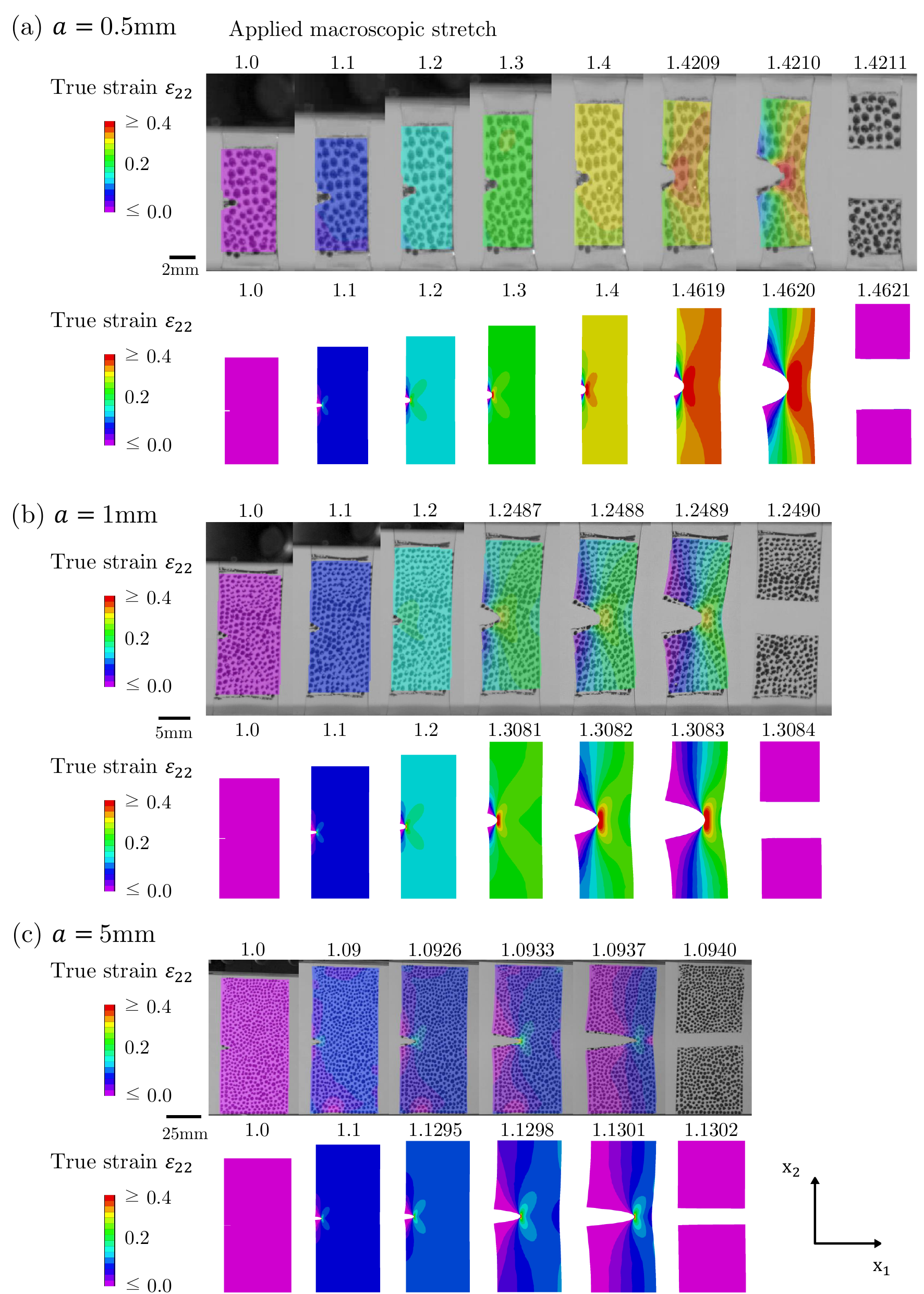} 
\caption{Progressive rupture of single-edge-notched PDMS specimens with the initial crack lengths of (a) $a$=0.5mm, (b) $a$=1mm, and (c) $a$=5mm in experiments and numerical simulations. Elements are removed in the simulation plots when the damage $d > 0.95$.}
\label{fig:simulation_result_PDMS}
\end{figure}

\begin{figure}[t!]
\centering
\includegraphics[width=0.61\textwidth]{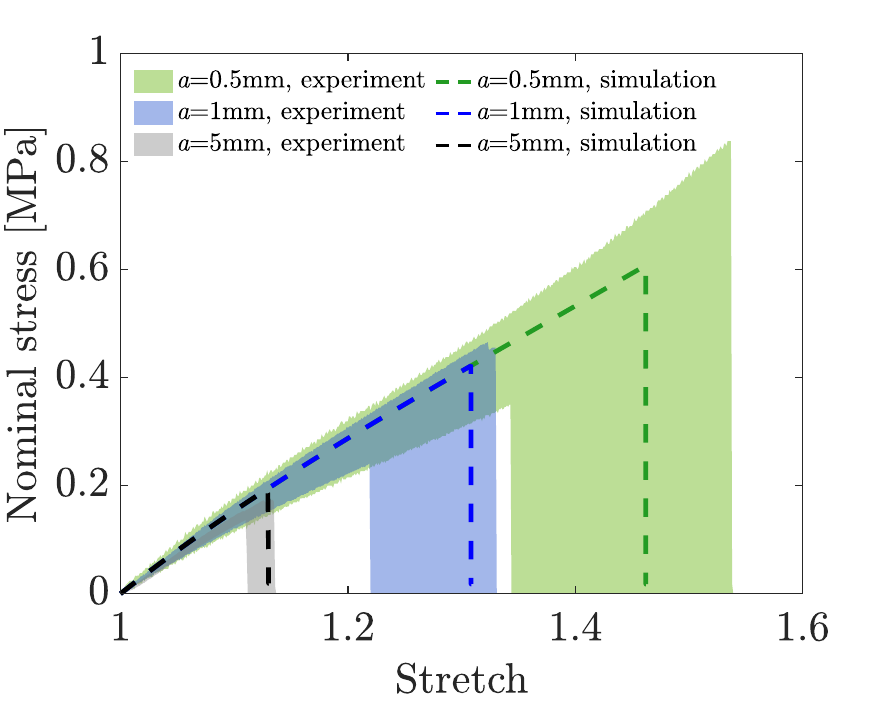} 
\caption{Nominal stress vs. stretch curves of single-edge-notched PDMS specimens with initial crack lengths $a=$ \{0.5, 1, 5\} mm in experiments (shaded lines) and numerical simulations (dashed lines). The shaded lines represent the experimental data of ten specimens.}
\label{fig:sim_stress-stretch_PDMS}
\end{figure}

Figure \ref{fig:simulation_result_PDMS} shows the sequential images of single-edge-notched specimens with various notch lengths of $a=$ \{0.5, 1, 5\} mm obtained from both experiments and numerical simulations. By making use of a digital image correlation (DIC) technique, strongly inhomogeneous axial true (logarithmic) strain fields in the vicinity of the notch were visualized during the entire fracture process (see Movie S1 available at \href{https://github.com/solidslabkaist/elastomer_fracture}{https://github.com/solidslabkaist/elastomer\_fracture}). Our numerical simulation results nicely captured the overall features and local (true) strain fields during initial crack extension and subsequent, progressive rupture to the final failure of the specimens into two parts. Furthermore, the final failure was found to be delayed in the smaller specimens in both experiments and numerical simulations. The experimental and numerical simulation results are further reduced into the macroscopic stress-stretch-failure curves in Figure \ref{fig:sim_stress-stretch_PDMS}. Though the specimens show the size-dependent fracture behavior, the initial stress-stretch responses in all specimens are almost identical due to the geometric similarity. The numerical simulations (dashed lines) are compared against the experimentally measured stress-stretch-failure curves (color shaded lines), yielding favorable agreement in each case of the PDMS specimens with the notch lengths of $a=$ \{0.5, 1, 5\} mm.

\subsubsection{Notch length-sensitivity governed by intrinsic length scale}
\label{section:PDMS_vs_TP}
We further investigate the notch length-sensitivity in elastomers with different intrinsic length scales. The notch lengths of the PDMS specimens presented in Section \ref{section:PDMS} were much larger than the intrinsic length scale $l$ ($a/l > 10$). Here, we examine the size-dependent fracture behavior of a TangoPlus whose intrinsic length scale is comparable to the notch length; i.e., $a/l \sim \mathcal{O}(1)$
\footnote{A microstructural inhomogeneity resulting from photopolymerization during 3D printing of the TangoPlus specimens was found to not impact the macroscopic fracture behavior. We further show the printing direction-independent fracture behavior in the TangoPlus material in the Appendix \ref{appendix:parameter}.}.
\begin{figure}[p!]
\centering
\vspace{-0.5in}
\includegraphics[width=0.83\textwidth]{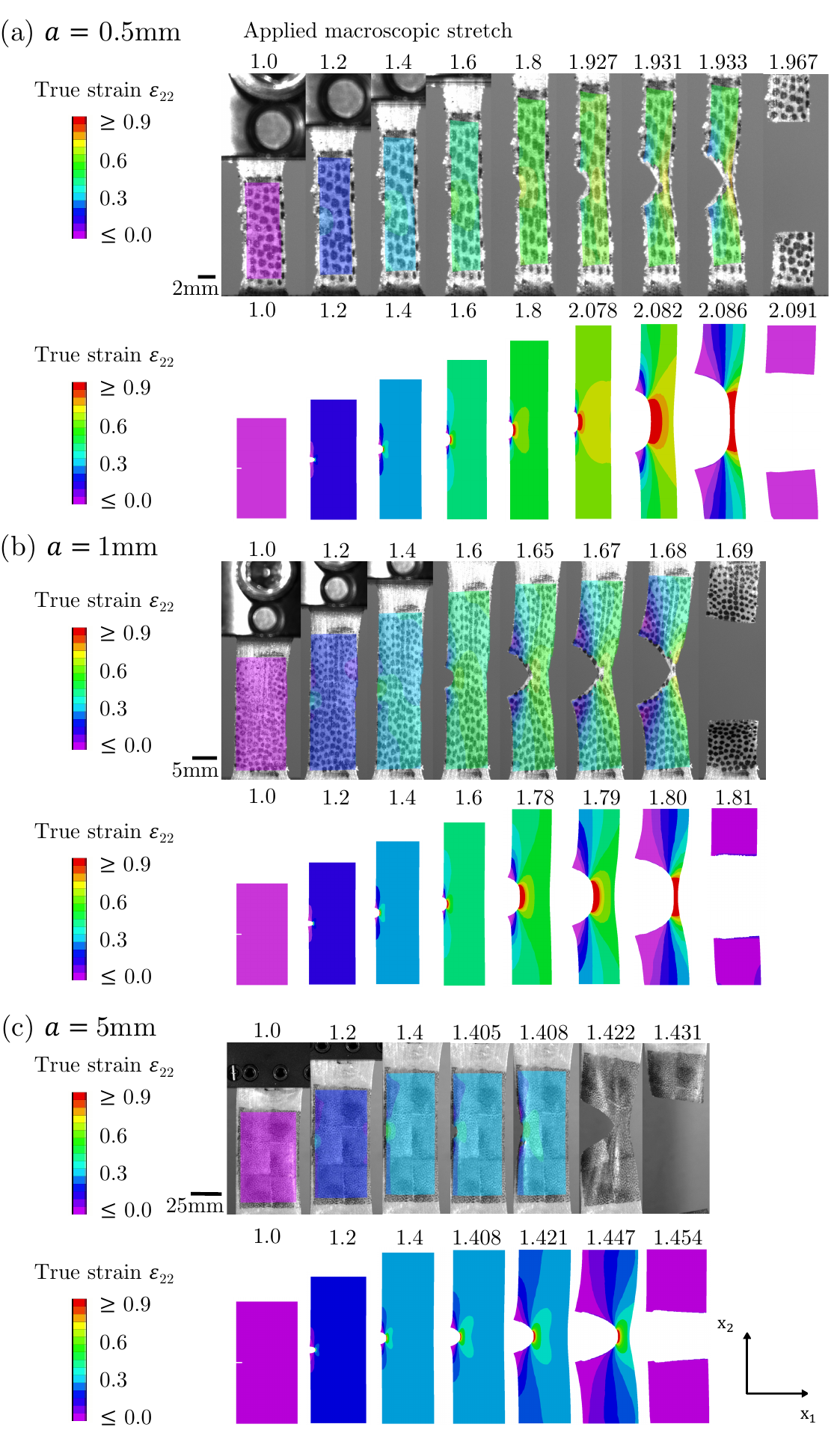} 
\caption{Progressive rupture of single-edge-notched TangoPlus specimens with the initial crack lengths of (a) $a$=0.5mm, (b) $a$=1mm, and (c) $a$=5mm in experiments and numerical simulations. Elements are removed in the simulation plots when the damage $d > 0.95$.}
\label{fig:simulation_result_TP}
\end{figure}

\begin{figure}[t!]
    \centering
    \includegraphics[width=0.65\textwidth]{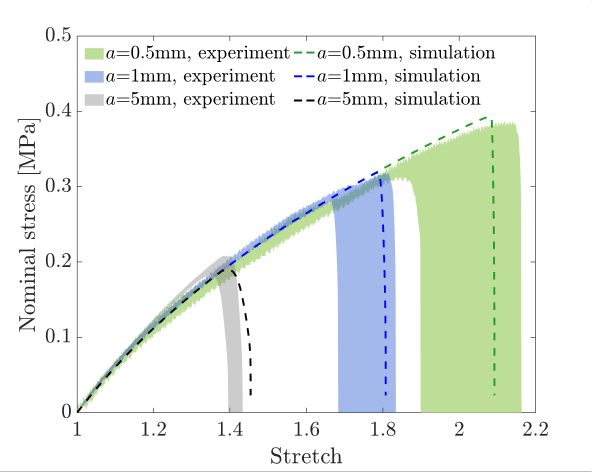}
    \caption{Nominal stress vs. stretch curves of single-edge-notched TangoPlus specimens with initial crack lengths $a=$ \{0.5, 1, 5\} mm in experiments (shaded lines) and numerical simulations (dashed lines). The shaded lines represent the experimental data of five specimens.}
    \label{fig:sim_stress-stretch_TP}
\end{figure}

Figure \ref{fig:simulation_result_TP} shows the experimental results of single-edge-notched TangoPlus specimens with notch lengths of $a=\{0.5, 1, 5\}$mm together with numerical simulations utilizing the experimentally identified intrinsic length scale $l$=1mm (see Movie S2 available at \href{https://github.com/solidslabkaist/elastomer_fracture}{https://github.com/solids labkaist/elastomer\_fracture}). The TangoPlus specimens have the same size as the PDMS specimens; we note that, in the TangoPlus specimens, a fixed ratio of notch-root radius to notch length, $R/a=0.1$ was used for geometric similarity\footnote{The average rupture stretches in the TangoPlus specimens with the fixed ratio of $R/a=0.1$ were found to be very similar to those of TangoPlus specimens notched using a razor blade ($R/a \ll 0.1$), supporting that the cracks made by 3D-printing were sufficiently sharp.}. The material parameters used in the simulations are summarized in Table \ref{Tab:tangoplus}. Detailed identification procedures for the material parameters are presented in the Appendix \ref{appendix:parameter}. The notch length-dependent fracture behavior of the TangoPlus specimens is clearly observed in both experiments and numerical simulations; the numerical simulations captured the local (true) strain fields during the entire fracture processes reasonably well. 
Furthermore, as shown in Figure \ref{fig:sim_stress-stretch_TP}, the numerical simulations nicely predicted the macroscopic stress-stretch-failure curves in each case of the TangoPlus specimens with the notch lengths of $a=\{0.5, 1, 5\}$ mm. Shaded lines represent the experimental data of five specimens.

\begin{table}[t!]
\doublespacing
\hspace{-0.1in}
\begin{tabular}{ccccccc}
\hline
 $\mu=N k_{b} \vartheta$ {[}MPa{]} & $K$ {[}MPa{]} & $\quad n \quad$ & $\varepsilon^{f}_{\mathrm{R}}$ {[}mJ/mm$^3${]} & $\bar{E}_b=NnE_{b}$ {[}MPa{]} & $l$ {[}mm{]} & $\zeta_{\mathrm{R}}$ {[}MPa$\,\cdot\,$s{]} \\
\hline
 0.24 & 24 & 10 & 0.18 & 10 & 1 & 0.5 \\
\hline
\end{tabular}
\vspace{-0.2in}
\caption{Material parameters used in the numerical simulations with the TangoPlus material.}
\label{Tab:tangoplus}
\end{table}

\begin{figure}[b!]
\hspace{-0.3in}
\includegraphics[width=1.1\textwidth]{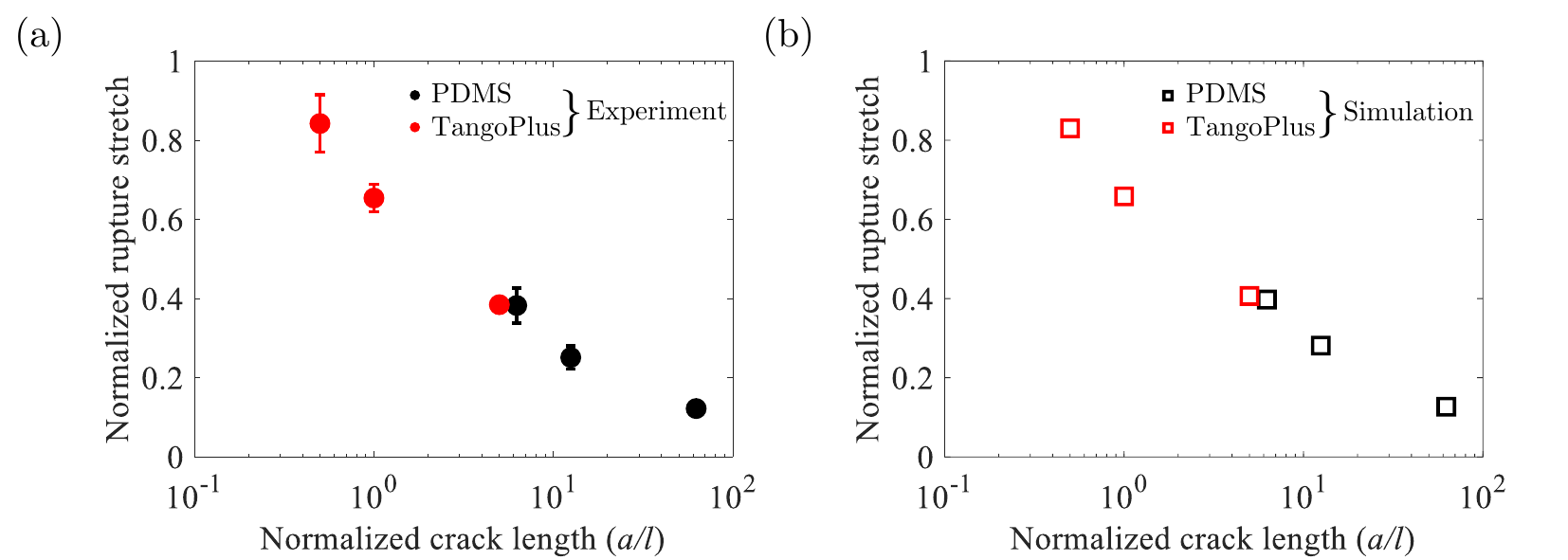} 
\caption{Rupture stretch (normalized by the rupture stretch in an unnotched specimen) vs. initial crack length (normalized by the intrinsic length scale) in PDMS and TangoPlus specimens in (a) experiments and (b) numerical simulations.}
\label{fig:rupture-stretch}
\end{figure}

Next, we present a quantitative comparison of the notch length-sensitivity that strongly depends on the intrinsic length scale in the two elastomers (here, PDMS and TangoPlus). To this end, we further reduced the stress-stretch-failure data from both experiments and numerical simulations to the rupture stretch normalized by the rupture stretch of an unnotched specimen ($2.60 \pm 0.20$ for PDMS and $2.34 \pm 0.10$ for TangoPlus; see Figure \ref{fig:parameter_PDMS} and \ref{fig:printing_direction} in the Appendix \ref{appendix:parameter}) against the initial crack lengths normalized by the intrinsic length scale ($l$ = 0.08 mm for PDMS and $l$ = 1 mm for TangoPlus) in Figure \ref{fig:rupture-stretch}. Note that both materials exhibited a comparable rupture stretch in unnotched specimens. However, the decrease in the rupture stretch (normalized by that in unnotched specimens) with an increasing notch length was found to be much greater in the TangoPlus specimens (0.85 to 0.39) than in the PDMS specimens (0.38 to 0.12) in both experiments (Figure \ref{fig:rupture-stretch}(a)) and numerical simulations (Figure \ref{fig:rupture-stretch}(b)). 

The higher notch-length sensitivity observed in the TangoPlus specimens reveals that the size-dependent fracture becomes more apparent when the initial crack length becomes closer to the intrinsic length scale; i.e., $a/l: 0.5 \sim 5$ in the TangoPlus specimens while $a/l: 5 \sim 50$ in the PDMS samples in our experiments. We further note that the TangoPlus specimen with an initial crack retains high stretchability comparable to that in the unnotched specimen ($a/l \ll 1$) though it exhibits the marked notch length-sensitivity, by which a slight increase in the notch length dramatically lowers the stretchability. In contrast, the PDMS specimen exhibits less sensitivity to the notch length but the stretchability remarkably lessens as against the unnotched specimen since the notch size is much larger than the intrinsic length scale in each of the notched specimens.
The notch length-sensitivity dependent on the ratio of the initial crack length to the intrinsic length scale ($a/l$) is attributed to the size of a “finite” fracture process zone as discussed in Section \ref{section:Fracture_process_zone}; i.e., when the notch length is significantly larger than the intrinsic length scale, the fracture process zone occupies a negligible region in the specimen. Hence, a change in the initial notch length has a slight impact on the fracture behavior. The nonlocal, gradient-damage model shows an excellent capability of describing these key features of the notch length-sensitivity in both elastomers.

Furthermore, the fracture process zone characterized by the intrinsic length scale significantly affects the crack-tip configuration during crack propagation throughout the specimens. As shown in Figures \ref{fig:simulation_result_PDMS} and \ref{fig:simulation_result_TP}, a “blunt” crack-tip geometry was observed in the TangoPlus in contrast to a sharp crack in the PDMS specimens with all notch lengths of $a=$ \{0.5, 1, 5\} mm. Our numerical simulations qualitatively captured the characteristics of the crack-tip geometry in the two elastomers with the different intrinsic length scale.

\subsubsection{Highly stretchable elastomers: PU and VHB}
\noindent
The predictive capability of the nonlocal continuum model utilizing the physically motivated intrinsic length scale is further examined by comparing the modeling results to the experimental data on highly stretchable materials, specifically commercially available synthetic elastomers, a polyurethane (PU) and an acrylate elastomer (VHB) recently reported in \cite{chen2017flaw}. Final failure in these elastomers occurs at much larger macroscopic stretches ($\sim$ 500\% and $\sim$ 1000\%), than in the TangoPlus ($\sim$ 100\% in stretch) and the PDMS ($\sim$ 150\% in stretch) cases. The material parameters used in the numerical simulations for the PU and VHB materials are summarized in Table \ref{Tab:pu_vhb}.

\begin{figure}[h!]
\centering
\includegraphics[width=1.0\textwidth]{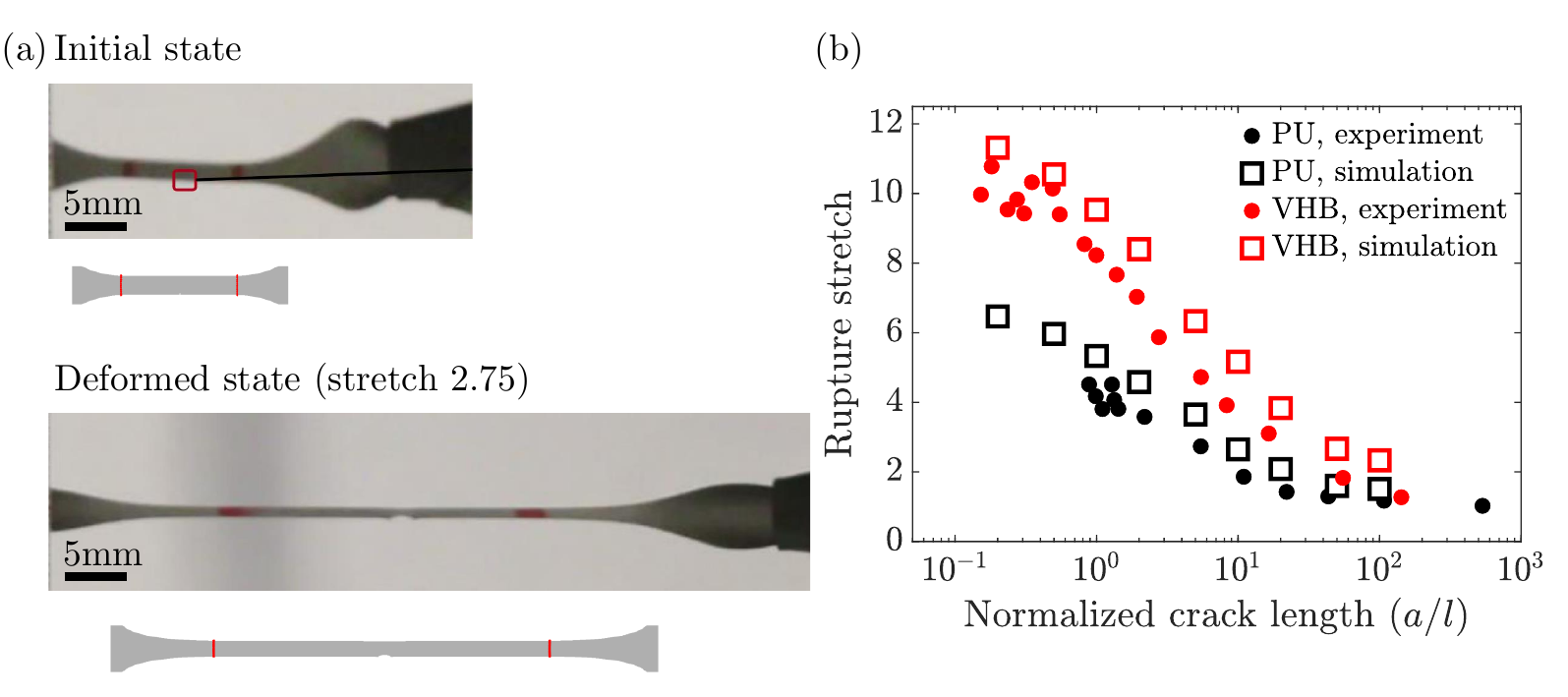} 
\vspace{-0.3in}
\caption{(a) Initial and deformed configurations of the PU specimen in experiments (reported in \cite{chen2017flaw}) and numerical simulations, and (b) rupture stretch vs. normalized crack length for the PU and VHB specimens in experiments (closed symbols; taken from \cite{chen2017flaw}) and numerical simulations (open symbols).}
\label{fig:ZSuo_PU_VHB}
\end{figure}

Figure \ref{fig:ZSuo_PU_VHB} displays the numerical simulation results (open symbols) of single-edge-notched PU and VHB specimens along with the experimental data (closed symbols; taken from \cite{chen2017flaw}) at a wide range of notch sizes from 0.1mm to 50mm. Additionally, the undeformed and deformed specimen configurations in the experiments and numerical simulations are presented for a visual aid. As shown in Figure \ref{fig:ZSuo_PU_VHB}(b), highly size-dependent fracture behavior is clearly evidenced through the macroscopic rupture stretch as a function of the initial crack length normalized by the intrinsic length scale, here $l$ = 0.093 mm for the PU elastomer ($l$ = 0.37 mm for the VHB elastomer) in both experiments and numerical simulations. As the length of the initial crack (or specimen sizes) approaches the experimentally identified intrinsic length scale (i.e., $a/l \sim$ 1), the size-dependence of the final rupture becomes more prominent. As shown, our numerical simulation results demonstrate an excellent ability to predict the size-dependent fracture behavior in these extremely stretchable elastomeric materials ($>$ 1000\% in stretch) over a wide range of initial crack lengths.

\begin{table}[h]
\doublespacing
\hspace{-0.4in}
\begin{tabular}{l|ccccccc}
\hline
& $\mu$ {[}MPa{]} & $K$ {[}MPa{]} & n & $\varepsilon^{f}_{\mathrm{R}}$ {[}mJ/mm$^3${]} & $\bar{E}_b$ {[}MPa{]} & $l$ {[}mm{]} & $\zeta_{\mathrm{R}}$ {[}MPa$\,\cdot\,$s{]} \\
\hline
 Polyurethane (PU) & 3.5 & 350 & 8 & 144.3 & 1500 & 0.093 & 0.5 \\
\hline
 Acrylate elastomer (VHB) & 0.028 & 2.8 & 20 & 5.58 & 60 & 0.37 & 0.5 \\
\hline
\end{tabular}
\vspace{-0.2in}
\caption{Material parameters used in numerical simulations for the PU and VHB materials}
\vspace{0.3in}
\label{Tab:pu_vhb}
\end{table}

\section{Application to complex geometries}
\label{section3}
\subsection{Notch-root radius sensitivity}
We explore in more depth how the notch geometry influences the fracture responses in elastomeric specimens of different sizes. We focus our attention here on the effects of the notch-root radius. The notch-root radius has been found to impact fracture responses significantly; it is also well known that the stress concentration near a sharp crack decreases the resistance against crack extension. However, recent studies have revealed that the stress concentration near the crack tip is not prominent at small length scales in amorphous materials (\cite{gao2003materials, pan2015origin, sha2019notch}). Here, we analyze the notch-root radius sensitivity strongly associated with the size effect via the intrinsic length scale $l$. To this end, we investigated the fracture response of TangoPlus specimens with various notch-rood radii, $R/a=$ \{0.1, 0.5, 2.5, 12.5\} where the initial notch length was taken to be $a=$ \{0.5, 1, 5\} mm. We further note that the specimen width was $w = 10a$ and the height was $h = 20a$; the specimen dimensions are the same as in the TangoPlus experiments presented in Section \ref{section2}\footnote{We have selected the TangoPlus material for analysis since the specimens with complex geometries can be easily fabricated.}$^,$\footnote{The printer resolution is comparable to the notch-root radius of the smallest specimens with notch length of $a=0.5$ mm, where $R/a = 0.1$. We note that the average rupture stretches for the specimens with $R/a=0.1$ ($a=0.5$ mm and 1 mm) are found to be comparable to those in specimens notched using a razor blade ($R/a \ll 0.1$), as discussed in Section \ref{section:PDMS_vs_TP}; i.e., the specimens with $R/a=0.1$ ($a = 0.5$ mm and 1 mm) are shown to have sufficiently sharp cracks.}.

\begin{figure}[t!]
\centering
\includegraphics[width=1.0\textwidth]{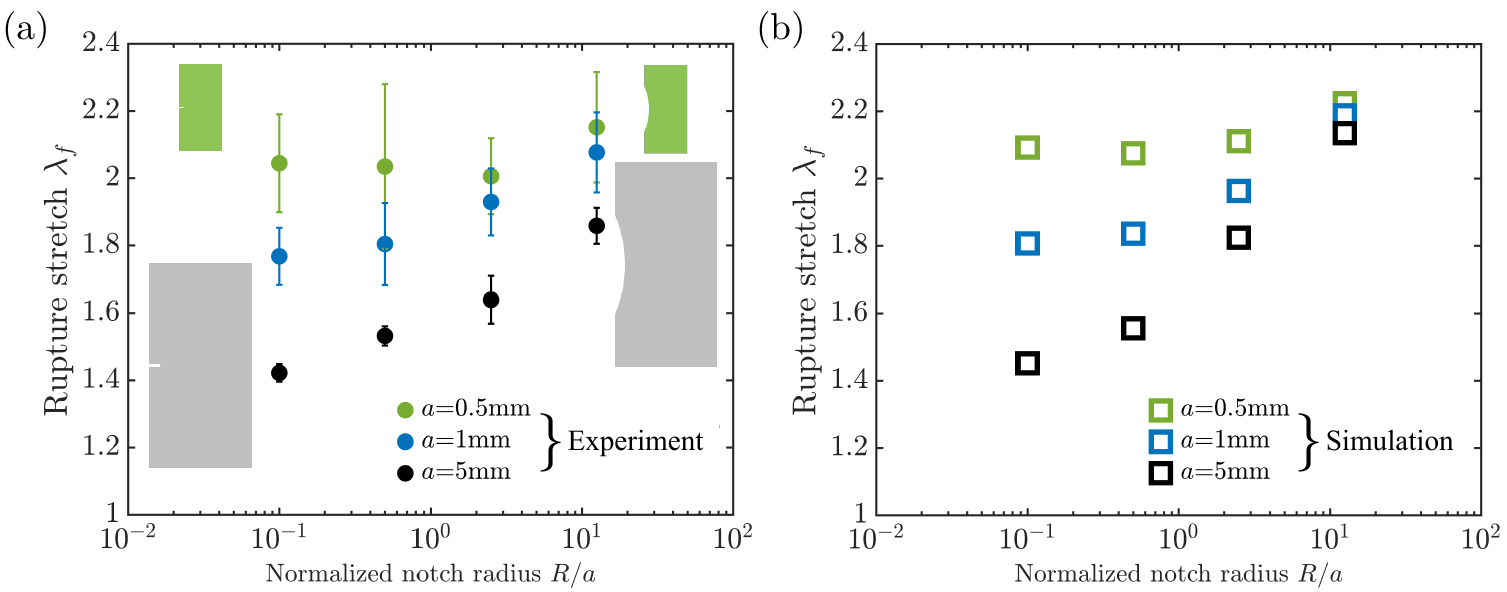} 
\caption{Rupture stretch vs. normalized notch-root radius of the single-edge-notched TangoPlus specimens in (a) experiments and (b) numerical simulations.}
\label{fig:radius_exp_sim}
\end{figure}

Figure \ref{fig:radius_exp_sim} shows the experimental and numerical results of rupture stretches as a function of the normalized notch-root radius with varying initial notch length (or specimen size). As clearly displayed in the experimental results in Figure \ref{fig:radius_exp_sim}(a), the macroscopic rupture stretch decreases significantly with a decrease in the notch-root radius due to the stress concentration near the notch tip, especially in the specimens with the largest notch length of $a =$ 5 mm (largest specimen size, black closed symbols). The notch-root radius-sensitivity lessens as the notch length decreases; i.e., the macroscopic rupture stretch is found to become insensitive to the notch-root radius when $R/a < 0.5$ in the specimens with notch length of $a=1$ mm (blue closed symbols). However, the specimens with the smallest notch length of $a=0.5$ mm (green closed symbols) does not exhibit the notch-root radius-sensitivity for any of $R/a =$ 0.1, 0.5, 2.5, and 12.5. The transition from notch-root radius-sensitive to -insensitive fracture behavior emanates from the presence of the fracture process zone near the notch tip, where the stress concentration is significantly mitigated. The size-dependent notch-root radius sensitivity in the fracture response is precisely captured in the nonlocal continuum model that makes use of the intrinsic length scale, as shown in Figure \ref{fig:radius_exp_sim}(b).

\subsection{Progressive rupture in a specimen randomly perforated}
\begin{figure}[b!]
\hspace{-0.4in}
\includegraphics[width=1.1\textwidth]{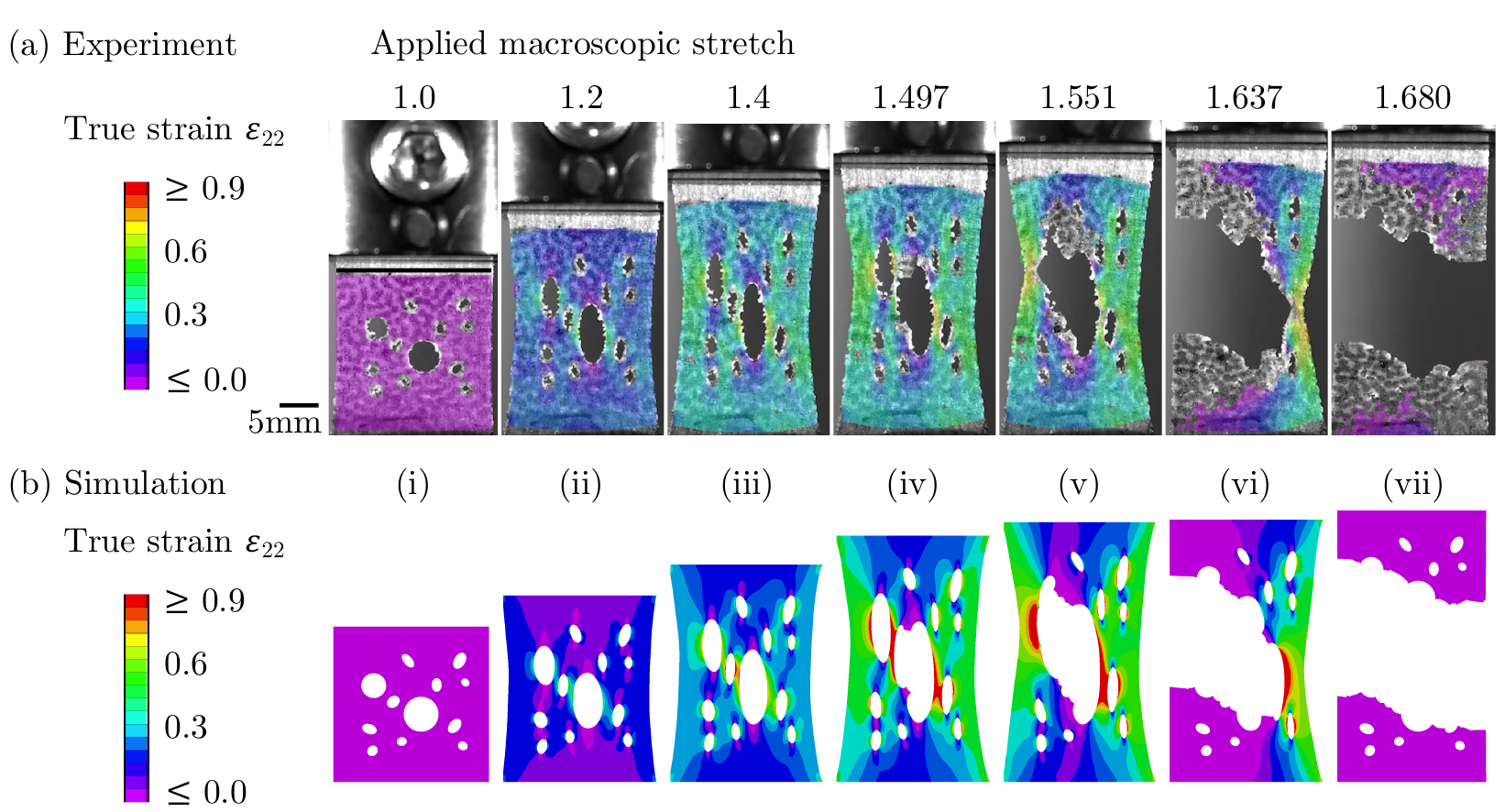} 
\vspace{-0.2in}
\caption{Progressive rupture of a TangoPlus specimen with randomly distributed holes in (a) experiment and (b) numerical simulation. Elements are removed in the simulation plots when the damage $d > 0.95$.}
\label{fig:random_sequential}
\end{figure}
\begin{figure}[t!]
\centering
\includegraphics[width=0.57\textwidth]{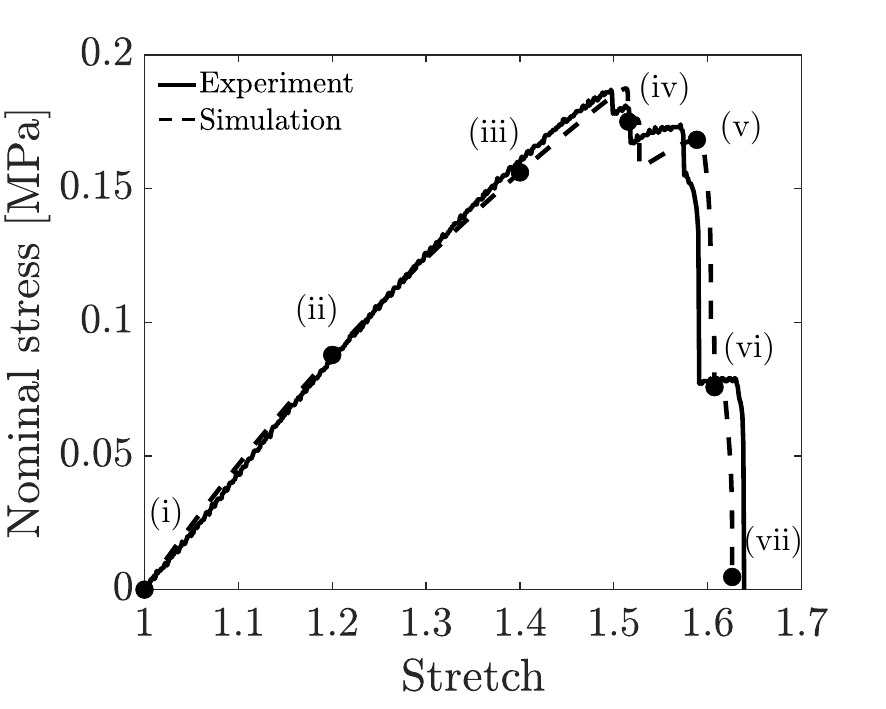} 
\caption{Nominal stress vs. stretch curves of the TangoPlus specimen with randomly distributed holes in experiment and numerical simulation.}
\vspace{0.3in}
\label{fig:random_stress_stretch}
\end{figure}

Next, we demonstrate the predictive capability of the nonlocal continuum model for a complicated fracture process involving the nucleation, propagation, and merging of cracks in an elastomeric specimen with an arbitrary complex geometry. Specifically, we explored a progressive rupture in a TangoPlus specimen with randomly distributed circular and elliptical holes under macroscopic tensile loading. The specimen size was 20 mm$\times$20 mm, and the thickness was 0.5 mm. Figures \ref{fig:random_sequential}(a) and (b) show sequential images of the specimen obtained from both experiments and numerical simulations, respectively, along with the macroscopic stress-stretch-failure curves in Figure \ref{fig:random_stress_stretch} (see Movie S3 available at \href{https://github.com/solidslabkaist/elastomer_fracture}{https://github.com/solidslabkaist/elastomer\_ fracture}). We further note that the material parameters used here are identical to those in the numerical simulations of single-edge-notched specimens summarized in Table \ref{Tab:tangoplus}. In the initial stages of tensile loading (i)-(iii), the ligaments between the holes were significantly stretched as the macroscopically imposed stretch increased. Then, a progressive rupture began throughout the most stretched ligaments in stage (iv), followed by the coalescence of cavities within the specimen, resulting in the formation of a larger cavity in stage (v). Finally, the specimen was separated into two parts during stages (vi) and (vii). Our numerical simulation excellently matches the experimentally measured stress-stretch curve with sudden stress drops that depict the load-bearing capacity loss throughout the sequentially fractured ligament network in the specimen, as shown in Figure \ref{fig:random_stress_stretch}.

\section{Concluding remarks}
\label{section4}
In this work, we explored nonlocal fracture behaviors in elastomeric materials that strongly depend on the sizes of pre-existing flaws by means of experiments and numerical simulations. We have demonstrated that a nonlocal continuum model can successfully predict all salient features of size-dependent fracture behaviors quantitatively in elastomeric materials that exhibit a wide range of stretchability ($\sim$100\% to $\sim$1000\% in stretch) prior to the final rupture. Central to the predictive capability of the nonlocal continuum model is the physically motivated intrinsic length scale $l$ of the elastomeric materials. We have experimentally determined the intrinsic length scale by measuring the critical energy release rate $\mathit{\Gamma}$ and the critical energy density $W^{*}$, and directly utilized it in a nonlocal phase-field framework able to describe the size-dependent fracture responses in these elastomers with varying initial crack lengths. Furthermore, through experiments and numerical simulations, the size-dependent fracture has been found to become more prominent as the notch length (or specimen size) approaches the intrinsic length scale of the elastomers. Overall, we have clearly shown that the physically motivated intrinsic length scale $l$ is crucial for the predictive modeling of the nonlocal, size-dependent fracture responses.

We have also conducted a comprehensive investigation of the impact of notch geometries on the nonlocal, size-dependent fracture responses in the elastomers. A transition from notch-root radius-sensitive to -insensitive fracture behavior was experimentally observed as the notch length decreased. The stress concentration near sharp cracks (small notch-root radius) was found to be effectively mitigated, especially in the smaller specimens where the fracture process zone is predominant over the entire region throughout the specimens. The nonlocal continuum model that utilizes the intrinsic length scale $l$ was also found to precisely predict the notch-root radius effects including the transition. More interestingly, we have demonstrated the predictive capabilities of the nonlocal continuum model for a complicated failure response of an elastomeric specimen with randomly distributed elliptic and circular holes. We further note that an overall fracture process involving crack initiation and propagation, coalescence of cavities, and the final failure throughout the geometrically complex specimen was excellently described in the nonlocal continuum model without any adjustments of the material parameters used in simulating the geometrically simple single-edge-notched specimens. Hence, we expect the nonlocal continuum model with suitable identification of the intrinsic length scale $l$ to be applicable to other technologically relevant geometries from microscopic- to continuum levels.

Throughout this work, we have focused our attention on fracture due to the scission of molecular bonds throughout elastomeric networks (\cite{lake1967strength, mao2017rupture, mao2018fracture}) without any consideration of bulk dissipation contributions. However, recent studies have revealed that more complicated dissipation mechanisms, such as rate-dependent inelasticity and Mullins' effect (\cite{cho2013constitutive, cho2017deformation, lee2023polyurethane, cho2024large, mao2017large}), play a significant role in the macroscopic fracture responses in elastomeric materials (\cite{slootman2020quantifying, kubo2021dynamic}).
Furthermore, there have been extensive research efforts to enhance the fracture toughness of brittle elastomers through sacrificial bond breaking (\cite{gong2003double, ducrot2014toughening}), stress-induced crystallization (\cite{trabelsi2002stress, persson2005crack}), and fiber reinforcement (\cite{wang2019stretchable, mo2022tough}) strategies. By incorporating other length scales associated with these additional dissipation mechanisms (\cite{tauber2020microscopic, edwards2020molecular}), the nonlocal continuum model can be further developed to better account for the inelasticity mechanisms and their impact on the size-dependent fracture in elastomeric materials. Moreover, predictive modeling of the fracture in more realistic elastomeric networks with various topological features (\cite{arora2020fracture,danielsen2021molecular,wang2021mechanism}) as well as in architected network materials (\cite{cho2016engineering, cho2024large, shaikeea2022toughness, karapiperis2023prediction}) can also be another focus of the next steps.

\section*{Acknowledgements}
We gratefully acknowledge fruitful discussions with Lallit Anand (MIT). HC acknowledges financial support from the National Research Foundation of Korea under grant number (RS-2023-00279843).

\renewcommand*\appendixpagename{Appendix}
\renewcommand*\appendixtocname{Appendix}
\begin{appendices}
\numberwithin{equation}{section}
\numberwithin{figure}{section}

\section{Experimental set-up}
A PDMS base material and a curing agent (SYLGARD 184, Dow Corning) were mixed at a weight ratio of 10:1. Air bubbles were removed by a vacuum desiccator. The mixture was then bar-coated on a flat, smooth glass substrate using a 0.5 mm-thick spacer. After 48 hours of curing at room temperature, followed by 1 hour at 100$^{\circ}$C, the thin PDMS sheets were detached from the glass substrate. The TangoPlus specimens were fabricated using a high-precision 3D printer (Connex3 Objet260, Stratasys Inc). The specimens were kept at ambient conditions for 24 hours. Then we conducted uniaxial tension tests for the PDMS and TangoPlus specimens at a strain rate of 0.01 s$^{-1}$ using a universal mechanical testing machine (Instron, Load-cell: 100N) at room temperature (294K). During the quasi-static tension tests, we measured the local deformation field near the crack in each of the specimens by tracking randomly distributed speckle patterns using the digital image correlation technique. Furthermore, a high-speed camera was utilized for imaging the fracture processes in the PDMS specimens.

\section{Identification of material parameters}
\label{appendix:parameter}
Here, we present detailed procedures to identify the material parameters used in our numerical simulations for the PDMS, TangoPlus, PU and VHB materials in the uniaxial tensile loading to fracture presented in Section \ref{section2} and \ref{section3}.

\subsection*{PDMS}
\begin{figure}[b!]
\centering
\includegraphics[width=0.6\textwidth]{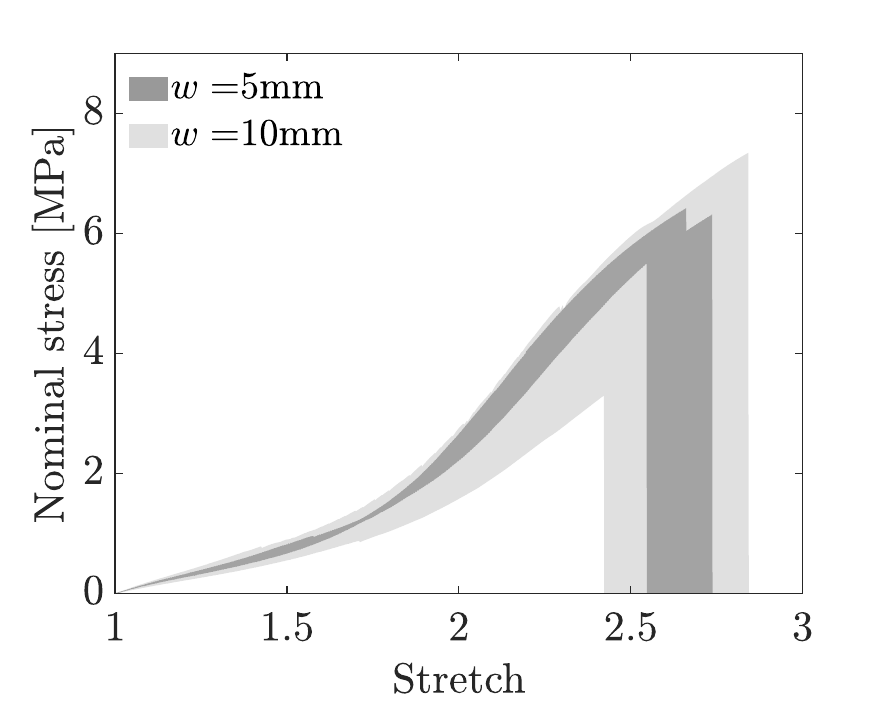} 
\vspace{-0.2in}
\caption{Nominal stress vs. stretch curves for unnotched specimens with a varying width of $w$ = 5 mm (dark shaded lines) and $w$ = 10 mm (light shaded lines). The shaded lines represent the experimental data of five specimens.}
\label{fig:parameter_PDMS}
\end{figure}

First, we determined the critical energy release rate $\mathit{\Gamma}$ and the critical energy density $W^{\ast}$ from the tension tests, by which we identified the intrinsic length scale $l=\mathit{\Gamma} / W^{\ast}$. We obtained the energy density $W(\lambda_{f})$ as a function of the macroscopic stretch to rupture $\lambda_{f}$ by measuring the work required for macroscopic failure in the single-edge-notched PDMS specimens with a wide range of initial crack lengths. Following the work of \cite{thomas1955rupture} based on the Griffith-type fracture criterion, the critical energy release rate was then obtained, $\mathit{\Gamma} \sim 0.25$ mJ/mm$^2$ from the relation $\mathit{\Gamma} = 2 \, a \, k(\lambda_{f}) \, W(\lambda_{f})$, where $a$ is the initial crack length and the non-dimensional factor, $k(\lambda_{f})$ is given by $\frac{3}{\sqrt{\lambda_{f}}}$ (\cite{greensmith1963rupture, lin2010large}). Furthermore, we obtained the critical energy density (or work to rupture) $W^{\ast}$ by computing the work required for rupture in unnotched specimens (\cite{chen2017flaw}). As shown in Figure \ref{fig:parameter_PDMS}, $W^{\ast}$ was found to be $\sim 2.7$ mJ/mm$^3$ from the stress-stretch curves of the unnotched PDMS specimens, where the work to rupture and the rupture stretch were approximately independent of the specimen size. The critical energy release rate $\mathit{\Gamma}$ and the critical energy density $W^{\ast}$ we obtained for the PDMS materials are very consistent with those reported in the prior studies (\cite{genesky2010toughness, johnston2014mechanical, wang2019stretchable}).

Next, the shear modulus and the limiting chain extensibility used in the modified Arruda-Boyce representation (see Equations (\ref{eq:entropyAB}) and (\ref{eq:stressAB})) were identified to be $\mu=0.38$ MPa and $\sqrt{n}=\sqrt{1.8}$ from the stress-stretch curves, also very consistent with those reported in the previous studies (\cite{ genesky2010toughness, johnston2014mechanical, wang2019stretchable}). Assuming the Poisson's ratio of $\nu =$ 0.495 (nearly incompressible behavior), the bulk modulus was taken to be $K=38$ MPa. Furthermore, the bond scission energy $\varepsilon^{f}_\mathrm{R} \sim$ 1.4 mJ/mm$^3$ in the governing equation of the damage field (see Equation (\ref{eq:strong_form})) was determined from the experimentally measured $W^{\ast}$. 
The bond stiffness per unit area was identified to be $\bar{E}_b =$ 30 MPa; note that the ratio, $\varepsilon^{f}_\mathrm{R} / \bar{E}_b \sim$ 0.06 we identified for our numerical simulations, was found to be very close to that estimated by the \textit{ab initio} calculations of the bond stretch reported in \cite{mao2017rupture}. Additionally, the kinetic parameter $\zeta_{\mathrm{R}}$ that controls the evolution rate of the scalar damage field in Equation (\ref{eq:strong_form}) was taken to be sufficiently small, approximately $\sim$ 1 kPa$\cdot$s, which does not critically influence the macroscopic failure response in our numerical simulations.

\subsection*{TangoPlus}
\begin{figure}[h]
    \centering
    \includegraphics[width=0.9\textwidth]{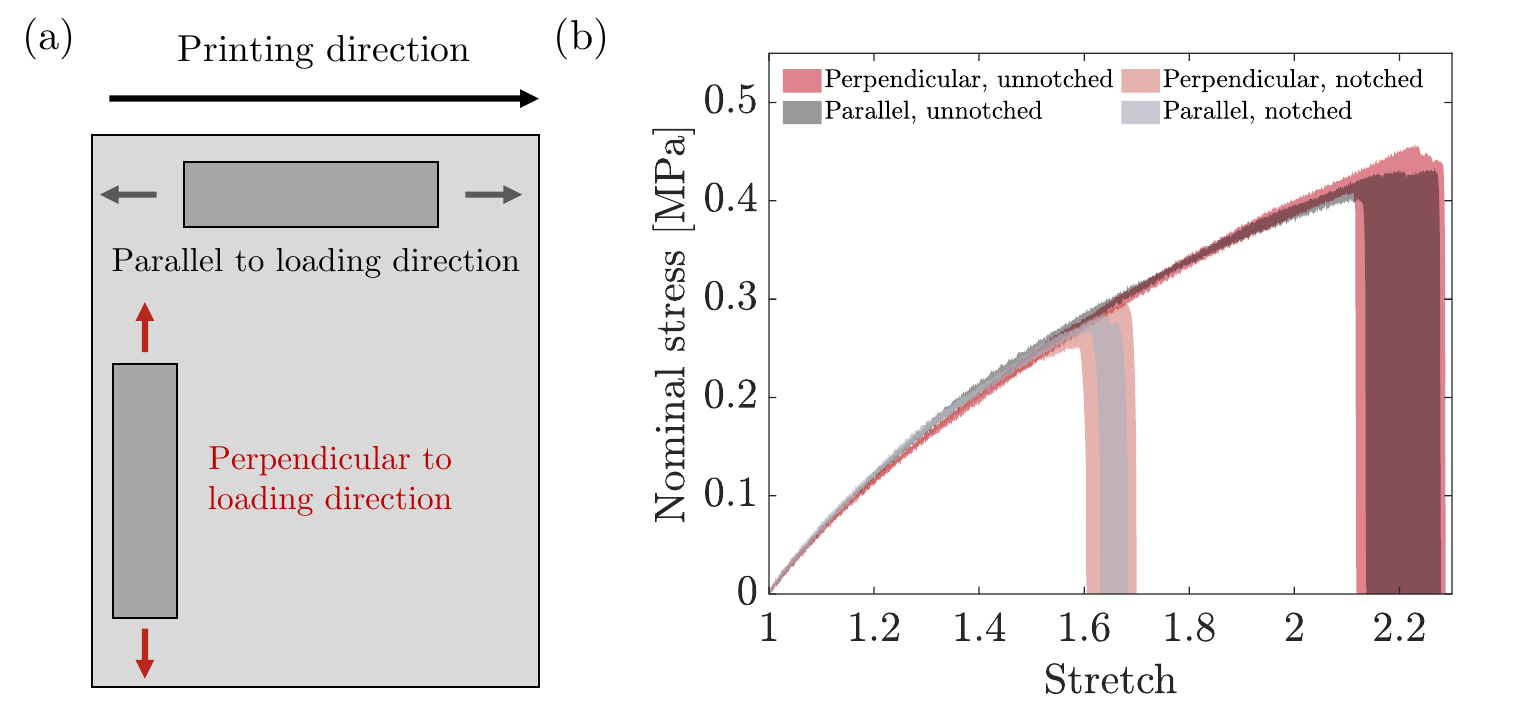}
    \caption{(a) A schematic of 3D-printed specimens and printing directions. (b) Stress-stretch curves of notched and unnotched TangoPlus specimens printed in two different directions. The shaded lines represent the experimental data of five specimens.}
    \label{fig:printing_direction}
\end{figure}

The material parameters for the TangoPlus elastomer were determined using the same procedure as in the PDMS elastomers. Note that the TangoPlus specimens inherently possess microstructural inhomogeneity incurred during the 3D-printing process. However, we found that the microstructural inhomogeneity did not affect the macroscopic stress-stretch-failure behavior critically. Specifically, we conducted tension tests on the TangoPlus specimens printed in two different directions: (1) parallel to the loading direction and (2) perpendicular to the loading direction as depicted in Figure \ref{fig:printing_direction}(a). Figure \ref{fig:printing_direction}(b) shows the stress-stretch-failure curves in unnotched specimens and single-edge-notched specimens printed in the two different directions, demonstrating nearly identical responses and rupture stretches independent of the printing direction. Furthermore, the printer resolution ($\sim$ 10 microns; provided by the manufacturer) is orders of magnitude smaller than the intrinsic length scale ($l \sim$ 1mm) and the specimen size ($w$: 5 $\sim$ 50 mm). We then identified the energy release rate $\mathit{\Gamma} \sim 0.5$ mJ/mm$^2$ and work to rupture $W^{\ast} \sim 0.45$ mJ/mm$^3$.

Note that the macroscopic rupture in the TangoPlus specimens occurred without any significant stress-stiffening behavior as evidenced in Figure \ref{fig:printing_direction}(b). This supports that the contribution of an internal energy change due to bond-stretch to the critical energy density for network failure is relatively small \footnote{See our prior work (\cite{lee2023finite}), on the driving force (entropy-driven vs. internal energy-driven) for failure in elastomeric networks.}. Hence, the bond dissociation energy $\varepsilon^{f}_\mathrm{R} \sim$ 0.18 mJ/mm$^{3}$ in Equation (\ref{eq:strong_form}) was taken to be much smaller than the experimentally determined $W^{\ast}$. The shear modulus $\mu$ and the limiting chain extensibility $\sqrt{n}$ were identified to be $\mu=0.24$ MPa and $\sqrt{n}= \sqrt{10}$ from the stress-stretch curves, very close to those reported in the previous studies (\cite{li2018instabilities, lei20183d}).


\subsection*{PU and VHB (data taken from \cite{chen2017flaw})}
\begin{figure}[h!]
\centering
\includegraphics[width=0.95\textwidth]{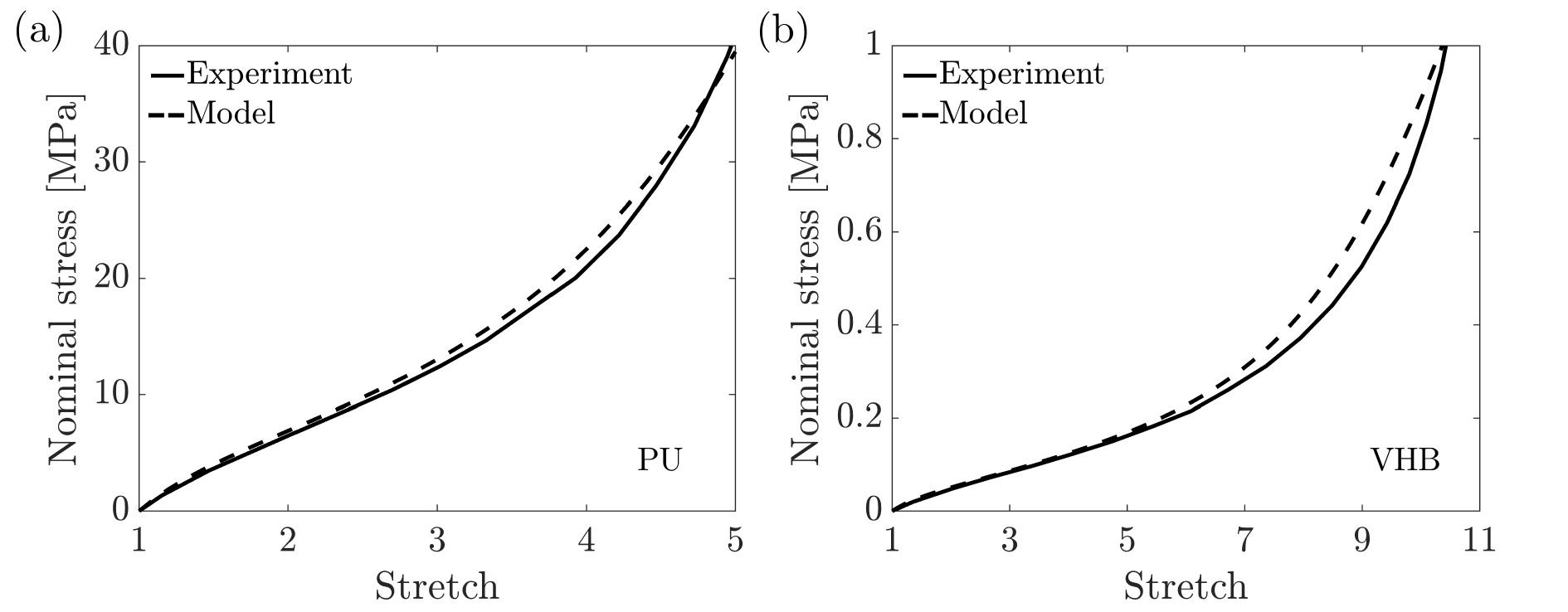} 
\caption{Nominal stress vs. stretch curves in uniaxial tension for (a) PU and (b) VHB materials (solid lines: experiments from \cite{chen2017flaw}, dashed lines: numerical simulations).}
 \label{fig:PU_VHB}
\end{figure}

The material parameters for PU and VHB were determined from the experimental data reported in \cite{chen2017flaw}. As shown in Figure \ref{fig:PU_VHB}, the shear modulus $\mu$ and the limiting chain extensibility $\sqrt{n}$ were obtained from the stress-stretch curves under uniaxial tension tests. Furthermore, we used $l =$ 0.093 mm for PU and $l =$ 0.37 mm for VHB reported in their paper, where the intrinsic length scale was identified from the experimentally measured $\mathit{\Gamma}$ and $W^{\ast}$. It should also be noted that $\varepsilon^{f}_{\mathrm{R}}$ in Equation (\ref{eq:strong_form}) was taken to be $\varepsilon^{f}_{\mathrm{R}} = W^{\ast} =$ 144.3 mJ/mm$^3$ for PU ($W^{\ast} = 5.58$ mJ/mm$^3$ for VHB) reported in their work; i.e., the internal energy change due to bond-stretch predominantly contributes to the work to rupture ($W^{\ast}$), as is well evidenced by the significant stress-stiffening behavior prior to final fracture in these highly stretchable elastomers.

\section{Numerical simulations}
The boundary value problems formulated in Equation (\ref{eq:strong_form}) were solved using a finite element procedure. The finite element procedure for the gradient-damage theory presented in Section \ref{section2} and \ref{section3} was implemented in a user-defined element (UEL) subroutine within Abaqus/Standard. Furthermore, a plane-stress condition ($T_{\mathrm{R}}(3,3) = 0$) was imposed (\cite{talamini2018progressive,lee2023finite}) throughout all numerical simulations. See our prior work (\cite{lee2023finite}) for numerical details on the finite element procedures. Abaqus input files and the UEL source codes are available online in the following GitHub repository, \href{https://github.com/solidslabkaist/elastomer_fracture}{https://github.com/solidslabkaist/elastomer\_fracture}.

\end{appendices}
\vspace{0.5in}

\onehalfspace
\printbibliography
\end{document}